\documentclass[journal]{IEEEtran}
\usepackage{cite}

\ifCLASSINFOpdf
\else
\fi
\usepackage{amsmath}
\usepackage{algorithmic}
\usepackage{url}

\usepackage{epsfig}
\usepackage{amsthm} 
\usepackage{mathrsfs}   
\usepackage{amsfonts} 
\usepackage{multirow} 
\usepackage{graphicx}
\usepackage{subfigure}
\usepackage{algorithm}
\usepackage{algorithmic}
\usepackage{dcolumn}
\usepackage{booktabs}
\usepackage[usenames, dvipsnames]{color}
\definecolor{myblue}{rgb}{0.0, 0.0, 1}
\usepackage{dcolumn}
\newcolumntype{d}{D{.}{.}{2.2}}  

\begin{document}
%

\title{Event-based EV Charging Scheduling in A Microgrid of Buildings}
%
%
%


\author{Qilong~Huang,~\IEEEmembership{Member,~IEEE},
        Li~Yang,~\IEEEmembership{Member,~IEEE},
        Chen~Hou,~\IEEEmembership{Member,~IEEE},
        Zhiyong~Zeng,~\IEEEmembership{Member,~IEEE},
        and Yaowen~Qi,~\IEEEmembership{Student Member,~IEEE}
\thanks{{This work has bee accepted by IEEE Transactions on Transportation Electrification. Please refer to the final version in IEEE.}}
\thanks{Q. Huang, L. Yang and Y. Qi are with the School of Automation, Nanjing University of Science and Technology, Nanjing 210094, China (email: huangql@njust.edu.cn, yangli945@126.com, qiyaowen@njust.edu.cn).}
\thanks{Li. Yang is the corresponding author.}
}

\maketitle

\begin{abstract}
With the popularization of the electric vehicles (EVs), EV charging demand is becoming an important load in the building. Considering the mobility of EVs from building to building and their uncertain charging demand, it is of great practical interest to control the EV charging process in a microgrid of buildings to optimize the total operation cost while ensuring the transmission safety between the microgrid and the main grid. We consider this important problem in this paper and make the following contributions. First, we formulate this problem as a Markov decision process to capture the uncertain supply and EV charging demand in the microgrid of buildings. Besides reducing the total operation cost of buildings, the model also considers the power exchange limitation to ensure transmission safety. Second, this model is reformulated under event-based optimization framework to alleviate the impact of large state and action space. By appropriately defining the event and event-based action, the EV charging process can be optimized by searching a randomized parametric event-based control policy in the microgrid controller and implementing a selecting-to-charging rule in each building controller. Third, a constrained gradient-based policy optimization method with adjusting mechanism is proposed to iteratively find the optimal event-based control policy for EV charging demand in each building. Numerical experiments considering a microgrid of three buildings are conducted to analyze the structure and the performance of the event-based control policy for EV charging.
\end{abstract}


%
\IEEEpeerreviewmaketitle

\section{Introduction}
%
%
%
%
\IEEEPARstart{E}{lectric} vehicles (EVs) have attracted more and more attention in recent years due to its lower emission and less dependence on fossil fuel. In order to achieve carbon peak and neutrality goals, China has made great effort to encourage the EV popularization. For example, the EV adoption increases nearly nine-fold in 2020 comparing to 2015. In 2020, there are about 4.17 million EVs on the road with 3:1 vehicle pile ratio\cite{CATARC2020}. Although the EV popularization helps to alleviate the fossil-fuel crisis and environment pollution, it brings a new challenge to the operation of microgrid if there is no charging control for the increasing EVs\cite{das2020electric}.

As EVs are charged by connecting to the charging piles in the building, the building is the main infrastructure which interacts with EVs. The impact of EV charging on the building lies in two aspects. \textit{On one hand}, the charging profile of EVs will influence the energy operation of the building. The building energy operator needs to procure extra power in order to satisfy EV charging demand and achieve load balance. If no proper charging control policy is implemented, the energy operation cost of the building may increase which will be paid by the users in the end. \textit{On the other hand}, if we only consider the charging control of EVs which parked in the corresponding building, the charging actions of multiple buildings may be homogeneous which may bring a new load peak to the grid. This may make the aggregated load exceed the contract capacity considering a microgrid with multiple buildings. When this happens, it may congest the distribution feeders and transformers which may cause voltage fluctuations in the microgrid\cite{teng2016design}.

According to the UK national travel survey, EVs are usually parked in the building 96.5\% of the time\cite{RAC2012}. Therefore, EVs can be considered as mobile storage devices with short unavailable time. In this context, it is of great practical interest to schedule the EV charging in a microgrid of buildings to reduce the EV charging impact on the economic operation of the building while limiting the load peak impact to the main grid. This problem is non-trivial to solve due to the following difficulties:

\textit{First}, uncertainties in the supply and demand side of the buildings. The uncertainty in the supply side of the buildings comes from the uncertain generation of distributed renewable energy\cite{huang2016multi}. The uncertainty in the demand side of the buildings comes from the uncertain charging demand of EVs. The arrival time, the parking time and the required charging energy are all uncertain before EV begins to park and charge in the building\cite{fallah2020charge}. \textit{Second}, the large state space and action space introduced by the large number of EVs in the buildings. Currently there are usually dozens to hundreds of charging piles in the building\cite{xiaodi2016method}. If all these charging piles are occupied, the charging states and charging actions will be high-dimension and increase exponentially with the number of EVs. This makes it difficult to find an optimal charging control policy for this problem. \textit{Third}, the limitation on the aggregated charging power in a microgrid of buildings. As aforementioned, a new load peak may appear if all the buildings implement homogeneous charging control policy. Therefore, the power exchange of each building in the microgrid should also be considered during scheduling in order to avoid the increasing of the load peak. This further increases the difficulty of solving this problem.

In recent years, the EV charging control attracts much attention due to the popularization of EVs and its impact to the grid\cite{al2019review,kang2021mapping}. Various control methods have been established for EV charging scheduling. Based on the control model of these works, the control approaches can be mainly divided into two categories: the determined control approach and stochastic control approach.

In the determined charging control approach, the EV charging process is usually formulated as a determined programming model. In other words, it assumes the future EV charging demand is known a prior, such as arrival time, parking duration and required charging energy. Therefore, many traditional methods can be applied to search an optimized charging control policy, such as linear/quadratic programming method\cite{Sun2018Optimal}, mixed-integer programming\cite{Koufakis2020Offline}, heuristic approach\cite{Shaaban2019Joint}, model predictive control\cite{Yang2019Decentralized}, etc. These methods enjoy the high optimization efficiency due to the assumption of perfect information of uncertainty.

However, the assumption introduced above is hard to obtain and these uncertainties cannot be underestimated in practice. For example, if the charging control policy is derived based on the determined programming model by assuming some EVs are parked for charging, this policy may be sub-optimal or even infeasible when these EVs are absent due to the prediction error. Therefore, many researchers study the stochastic control method, such as scenario-based optimization\cite{fallah2020charge}, robust optimization\cite{Zhao2017Robust}, reinforcement learning\cite{LHP2020}, simulation-based policy improvement\cite{HQL2018Robust}, etc. The first two methods usually try to convert the stochastic charging control problem into a determined control problem. These two methods should be carried out periodically and there is no experience accumulation among the control policies obtained at each time. The latter two methods usually formulate the problem as a Markov decision process (MDP) and can be considered as a state-based method. Due to the well-known curse of dimensionality\cite{chang2007simulation}, these methods suffer the large state space and action space considering large number of EVs in a microgrid of buildings.

Most of existing works mainly consider the EV charging control in a parking lot or charging station\cite{Shaaban2019Joint,fallah2020charge}. The interaction between EV charging and building energy operation also gains attention in recent years. In \cite{Yang2019Decentralized}, it considers coordination between EV charging and building integrated wind energy. In \cite{elgqvist2019evaluating}, it studies the EV charging impact on the operation cost of the building. In \cite{wu2018transactive}, a transactive real-time EV charging management scheme is proposed to coordinate EV charging with the distributed photovoltaic (PV) generation in the building. However, few works consider the EV charging control in a microgrid of buildings to avoid homogeneous charging actions. This is critical to avoid exceeding the load capacity of the microgrid and ensure the operation safety.

Based on the discussions above, we study the EV charging scheduling problem in a microgrid of buildings in this paper. Compared with the published literature, the main contributions of this paper are as follows:

1)	The EV charging scheduling is formulated as a Markov decision process model to capture the uncertain distributed renewable energy and EV charging demand in the microgrid of buildings. Besides reducing the charging impact on the economic operation of the buildings, this model also limits the power exchange to the main grid to avoid the increasing of load peak for transmission safety.

2)	In order to alleviate the impact of large state space and action space when solving the proposed MDP model, we reformulate it within event-based optimization framework. By appropriately defining the event and event-based action, the EV charging process can be optimized by searching a randomized parametric event-based control policy in the microgrid controller and implementing a selecting-to-charging rule in each building controller.

3)	A constrained gradient-based policy optimzation method with adjusting mechanism is proposed to iteratively find the optimal event-based control policy for EV charging demand in each building while ensuring transmission safety. Numerical experiments considering a microgrid of three buildings are conducted to analyze the structure and the
performance of the event-based control policy for EV charging.

The rest of this paper is organized as follows. We formulate the problem in Section II, present the solution methodology in Section III, discuss the numerical results in Section IV, and briefly conclude in Section V.

\section{Problem Formulation}

\subsection{System Description}
\begin{figure}
\centering
\setlength{\abovecaptionskip}{-6pt}
\includegraphics[height=0.19\textheight]{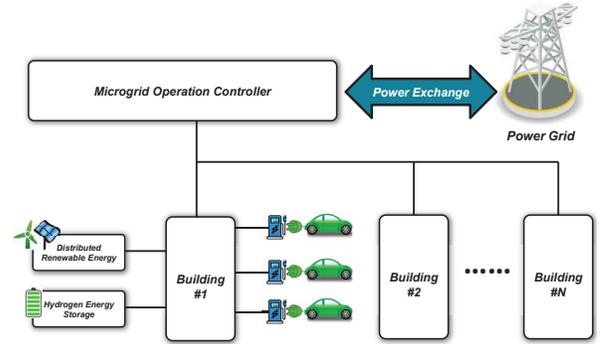}
\caption{System description of EV scheduling in a microgrid of buildings.}
\label{SystemOverview}
\end{figure}

We consider a microgrid of buildings as depicted in Fig. \ref{SystemOverview}. In the microgrid, each building is equipped with distributed renewable energy (DRE), hydrogen energy storage (HES) and charging piles. The building should provide charging service and keep load balance. We assume that only when the output of DRE and HES cannot satisfy the EV charging demand and building load, the building will procure power from the grid through microgrid operation controller. The building can also sell power to the grid if the output of DRE is large. In order to reduce the EV charging impact and ensure transmission safety, microgrid operation controller should regulate EV charging behaviors in each building based on the supply and demand status.

In the following, we will first formulate this multi-stage stochastic decision problem as Markov decision process to capture the uncertain renewable energy and EV charging demand in each building. To simplify the discussions, the following assumptions are made:

1)	The charge level of each EV is fixed.

2)	The distributed renewable energy is free.

3)  The electricity price from the grid is deterministic and known a priori.

\subsection{System Model}
We consider this EV scheduling problem in a microgrid of buildings over the discretized horizon $t=1,2,...,T$ where $t$ denotes the decision epoch and $\Delta T$ denotes the decision interval. There are $K$ buildings and $N$ EVs in the microgrid. The MDP model of the proposed problem is shown below.

\textit{1) System States:} The system state at stage $t$ is defined as $S_t=[s_t^1,s_t^2,...,s_t^K]$ where $k=1,2,...,K$ and $s_t^k$ denotes the state of the $k$th building. For each $s_t^k$, it is defined as $s_t^k=[r_t^k,b_t^k,n_{t,m}^k,n_{t,c}^k]$ where $r_t^k$ denotes the output of DRE in the building, $b_t^k$ denotes the State of Charge (SOC) of HES, $n_{t,m}^c$ denotes the number of EVs which must be charged at stage $t$ and $n_{t,c}^c$ denotes the number of EVs which can be charged at stage $t$.

\textit{2) Action Space:} The control action at stage $t$ is defined as $A_t=[a_t^1,a_t^2,...,a_t^K]$ where $a_t^k$ denotes the specific action for the $k$th building. Each building should decide whether to provide charge service for the connected EVs at stage $t$. Therefore, there is $a_t^k=[z_{t,1}^k,z_{t,2}^k,...,z_{t,N}^k]$ where $z_{t,i}^k \in \{0,1\}, i=1,2,...,N$. When $z_{t,i}^k=1$, it means the $k$th building should provide charge service for the $i$th EV at stage $t$ if it is parked, otherwise $z_{t,i}^k=0$.

\textit{3) System Dynamics:} As the energy status of EV and HES are both time-coupled, their system dynamics should be considered when action $A_t$ is decided for the current state $S_t$.

For each EV, we use a tuple $(T_t^i,E_t^i,L_t^i)$ to represent its remaining parking time, remaining required charging energy and parking location. $L_t^i \in \{0,1,2,...,K\}$ and $L_t^i = 0$ if the $i$th EV is on the road. Then, there is
\begin{equation}
\label{EQTimeDy}
T_{t+1}^i =
\begin{cases}
T_t^i \!-\! \Delta T,& \text{if} \ L_t^i > 0\\
\tau_{t+1}^i,& \text{if} \ L_{t+1}^i \times (1-L_{t}^i) > 0\\
0,& \text{if} \ L_{t+1}^i = 0
\end{cases}
\end{equation}
\begin{equation}
\label{EQEnerDy}
E_{t+1}^i =
\begin{cases}
E_t^i \!-\! z_t^i P \Delta T \psi^c ,&\text{if} \ L_t^i > 0\\
\eta_{t+1}^i,&\text{if} \ L_{t+1}^i \times (1-L_{t}^i) > 0\\
0,&\text{if} \ L_{t+1}^i = 0
\end{cases}
\end{equation}
where $P$ is the charge power, $\psi^c$ denotes the charge efficiency, $\tau_{t+1}^i$ and $\eta_{t+1}^i$ are both nonnegative random variables which denote the stochastic characteristic of EV charging demand in the future. As the location transitions for EVs are time-variant, the location transition probability can be established as
\begin{equation}
\label{EQEVLocationTrans}
P(L_{t+1}^i|L_{t}^i) =
\begin{bmatrix}
p_{11}(t) & p_{12}(t) & ... & p_{1K}(t) & p_{10}(t)\\
p_{21}(t) & p_{22}(t) & ... & p_{2K}(t) & p_{20}(t)\\
... & ... & ... & ... & ...\\
p_{K1}(t) & p_{K2}(t) & ... & p_{KK}(t) & p_{K0}(t)\\
\end{bmatrix}
\end{equation}
where $p_{K1}(t)$ denotes the EV is parked in the $K$the building at stage $t$ and moves to the first building, and so forth.

According to \cite{long2021efficient}, the system dynamics of HES in each building can be depicted as follows
\begin{equation}
\label{HESCapacityEq}
\kappa_{t+1}^k =
\begin{cases}
\max \{\kappa_{t}^k-h_t^k / \varphi^{\text{H2P}}, 0 \}, &\text{if} \ h_t^k \geq 0\\
\min \{ \kappa_{t}^k-h_t^k \varphi^{\text{P2H}}, \kappa^{\text{cap}} \}, &\text{if} \ h_t^k \leq 0
\end{cases}
\end{equation}
where $\kappa_{t}^k$ is the stored hydrogen of HES in the $k$th building at stage $t$, $\kappa_{\text{cap}}$ is the hydrogen storage capacity of HES, $\varphi^{\text{H2P}}$ is the round-trip efficiency from hydrogen to power, $\varphi^{\text{P2H}}$ is the round-trip efficiency from power to hydrogen, $h_t^k$ is the discharge power of HES if $h_t^k \geq 0$, otherwise is the charge power of HES. Considering $\kappa_{e,t}^k = \kappa_{t}^k \sigma_{H_2}$ where $\kappa_{e,t}^k$ is the stored hydrogen energy and $\sigma_{H_2}$ is the lower heating of hydrogen, equation (\ref{HESCapacityEq}) can be rewritten as follows by multiplying both sides with $\sigma_{H_2}/\kappa_{e}^{\text{cap}}$
\begin{equation}
b_{t+1}^k =
\begin{cases}
\max \{b_{t}^k-h_t^k / (\eta^{dc} \kappa_{e}^{\text{cap}}), 0 \}, &\text{if} \ h_t^k \geq 0\\
\min \{b_{t}^k-h_t^k \eta^{c} / \kappa_{e}^{\text{cap}}, 1 \}, &\text{if} \ h_t^k \leq 0
\end{cases}
\end{equation}
where $\kappa_{e}^{\text{cap}} = \kappa^{\text{cap}}\sigma_{H_2}$ denotes the energy capacity of HES, $\eta^{dc} = \varphi^{\text{H2P}}/\sigma_{H_2}$ denotes the discharge efficiency of HES and $\eta^{c} = \varphi^{\text{P2H}} \sigma_{H_2}$ denotes the charge efficiency of HES.

For each HES in the building, we regulate that HES will discharge when the DRE cannot satisfy the building demand and will charge if DRE is sufficient to meet the building demand, i.e.,
\begin{equation}
h_{t}^k =
\begin{cases}
-\min \{r_{t}^k-l_t^k - p_t^k, \overline{h_{t,c}^k} \}, &\text{if} \ r_t^k \geq l_t^k + p_t^k\\
\min \{l_t^k + p_t^k - r_{t}^k, \overline{h_{t,dc}^k} \}, &\text{if} \ r_t^k \leq l_t^k + p_t^k
\end{cases}
\end{equation}
where $l_t^k$ denotes the net demand in the $k$th building, $p_t^k = \sum_{i=1}^N z_{t,i}^kP$ is the total charge power in the $k$th building, $\overline{h_{t,c}^k}$ and $\overline{h_{t,dc}^k}$ satisfy
\begin{equation}
\begin{aligned}
\overline{h_{t,dc}^k} = \min \{h^{\text{cap}},b_t^k \kappa_{e}^{\text{cap}} \eta^{dc} \},  \\ \overline{h_{t,c}^k} = \min \{h^{\text{cap}},(1-b_t^k) \kappa_{e}^{\text{cap}} / \eta^{c} \}
\end{aligned}
\end{equation}
in which $h^{\text{cap}}$ is the maximum charge/discharge power of HES.

Based on the supply demand status in the building, each building can sell excess power or procure power from the grid. Therefore, the exchange power $g_t^k$ between the building and the grid can be depicted as follows,
\begin{equation}
g_{t}^k =
\begin{cases}
\max \{l_t^k + p_t^k - r_{t}^k - h_t^k, 0 \}, &\text{if} \ h_t^k \geq 0\\
-\max \{r_{t}^k + h_t^k - l_t^k - p_t^k, 0 \}, &\text{if} \ h_t^k < 0
\end{cases}
\end{equation}

\textit{4) Constraints:} The action $A_t$ corresponding to the state $S_t$ should satisfy the following constraints,
\begin{equation}
\label{EQTotalExchangePower}
\sum_{k=1}^K g_{t}^k = G_t
\end{equation}
\begin{equation}
\label{EQTotalExchangeBound}
\underline{G} \leq G_t \leq \overline{G}
\end{equation}
\begin{equation}
\label{EQExchangeBound}
\underline{g^k} \leq g_t^k \leq \overline{g^k}
\end{equation}
\begin{equation}
\label{EQEnergyBound}
0 \leq E_t^i \leq E_{\text{cap}}
\end{equation}
\begin{equation}
\label{EQEnerTimeCons}
0 \leq E_t^i \leq P \cdot B_t^i
\end{equation}
\begin{equation}
\label{EQLoadBalance}
r_t^k + h_t^k + g_t^k = l_t^k + p_t^k
\end{equation}
. Constraint (\ref{EQTotalExchangePower}) denotes the total exchange power in a microgrid of buildings. Note that constraints (\ref{EQTotalExchangeBound}) and (\ref{EQExchangeBound}) regulate the lower and upper bound of the total exchange power of the microgrid and the exchange power of the building to ensure transmission safety, respectively. Constraint (\ref{EQEnergyBound}) regulates that the required charging energy of each EV should not exceed the battery capacity. Constraint (\ref{EQEnerTimeCons}) constrains the remaining required charging energy should not exceed the maximum energy that can be supplied during the remaining parking time. This constraint is used to satisfy the driver's charging demand. Constraint (\ref{EQLoadBalance}) denotes the load balance in each building.

\textit{5) Objective Function:} As the responsibility of the microgrid operator controller is to optimize the operation cost during the entire periods, the following expected cumulative cost within a sliding window is chosen as the objective function considering the uncertain charging demand and output of DRE in the buildings, i.e.,
\begin{equation}
\label{EQObj}
\min J = \sum_{t=t_0}^{t_0 + T_w -1}\mathbf{E}^{\pi} [\sum_{k=1}^K c_t(s_t^k,a_t^k) ]
\end{equation}
where $t_0$ denotes the current decision stage, $T_w$ denotes the sliding window, $\pi$ denotes the EV scheduling policy and $c_t(s_t^k,a_t^k)$ denotes the one-step cost incurred by taking action $a_t^k$ at state $s_t^k$ for the $k$th building which is defined as follow:
\begin{equation}
\label{EQOneStepCost}
c_t(s_t^k,a_t^k) =
\begin{cases}
\omega_t (l_t^k + p_t^k + \overline{h_{t,c}^k} - r_{t}^k), \\
 \qquad  \qquad \text{if} \ p_t^k \leq r_{t}^k - l_t^k - \overline{h_{t,c}^k} \\
\omega_t (l_t^k + p_t^k - \overline{h_{t,dc}^k} - r_{t}^k), \\
\qquad \qquad  \text{if} \ p_t^k \geq r_{t}^k - l_t^k + \overline{h_{t,dc}^k} \\
0, \qquad \quad \text{else}
\end{cases}
\end{equation}
where $\omega_t$ denotes the electricity price.

Based on the proposed model above, the microgrid operator controller should minimize $J$ to find an optimal scheduling policy $\pi^*$ for each decision stage. However, due to the coupled constraint (\ref{EQTotalExchangeBound}) and the large state space and action space of the problem, the exact optimal solution of the above model can rarely be derived\cite{bertsekas2012dynamic}. In the next section, we will explore a event-based approach with gradient search to approximately solve the problem.

\section{Solution Methodology}
\subsection{Event Definition}
The proposed MDP model suffers large state space difficulty as it is a state-based model and its state space increases exponentially with the number of buildings and EVs. Therefore, we propose a event-based optimization (EBO) framework to solve the large state space difficulty. In the EBO framework, the model and the solution focus on the event which depicts the sets of state transition\cite{Cao2007Book}. The decision is event-triggered which can save computation burden. When the event is defined approximately, EBO can be applied to solve MDPs with large state space\cite{jia2018event}. Due to these advantages, EBO has been applied in various fields, such as HVAC control\cite{jia2018event,wu2015optimal}, communication network\cite{wardi2018perturbation}, stock trading\cite{xue2021optimization}, etc.

In this paper, our idea to use EBO comes from the fact that it may not need to describe the detailed EV charging state and output of the DRE which incurs the large state space. Instead, we can use event to depict the elastic ratio of the EV charging, DRE and HES. The larger elastic ratio means the larger control margin during scheduling.

Based on this idea, we can firstly define the elastic ratio of EV charging in each building, i.e.,
\begin{equation}
\label{EQEVElastic}
I_{t,EV} = \frac{(n_{t,c}^k - n_{t,m}^k)P}{(\overline{n^k}P)}
\end{equation}
where $\overline{n^k}$ denotes the number of charging piles in the $k$th building. Equation (\ref{EQEVElastic}) describes the elastic degree of charging which can be delayed. Secondly, the elastic ratio of HES in each building can be defined as its SOC, i.e., $I_{t,HES} = b_{t}^k$. The larger $b_{t}^k$ means that the HES can provide a more flexible charge/discharge power for operation cost optimization. Thirdly, the elastic ratio of DRE can be defined as follows,
\begin{equation}
\label{EQDREElastic}
I_{t,DRE} = \frac{r_t^k}{\overline{r^k}}
\end{equation}
where $\overline{r^k}$ denotes the generation capacity of the DRE in the $k$th building. Equation (\ref{EQDREElastic}) describes the generation level of the DRE and the higher level indicates the larger potential of operation cost saving.

Based on the elastic ratios introduced above, we can finally define the event as follows,
\begin{equation}
\label{EQEvent}
\mathcal{E} =\{ e_t^k | t=1,2,...,T, k=1,2,...,K\}
\end{equation}
where
\begin{equation}
e_t^k = \{ <s_{t-1}^k,s_{t}^k> | (I_{t,EV} + I_{t,HES} + I_{t,DRE})/3 \in [0,1]\}
\end{equation}
. In equation (\ref{EQEvent}), $e_t^k$ denotes the triggered event in the $k$th building at stage $t$. The value of $e_t^k$ is within $[0,1]$. If we divide this interval equally with discrete unit set as 0.1, the event space for each building is limited as 10 which is largely reduced compared with the large state space in the proposed MDP model.

\subsection{Randomized Parametric Event-based Policy}
Another difficulty of the proposed MDP is the large action space. It is of great computation burden to compute the charge control variables $z_{t,i}^k$ for each EVs. Therefore, we propose a randomized parametric event-based policy to alleviate the large action space impact of this problem.

The charge control of each EV is implemented into two steps. Firstly, the microgrid operator controller decides a parametric charge ratio $\alpha_t^k$ for each building and is chosen as the event-based action, i.e., $a_t^k = \alpha_t^k \in [0,1]$. In this way, the total charge power for each building can be described as follows,
\begin{equation}
p_t^k = [n_{t,m}^k + a_t^k (n_{t,c}^k - n_{t,m}^k)]P
\end{equation}
. As the charge ratio $\alpha_t^k \in [0,1]$, the action space of the proposed problem can be largely restricted. Based on \cite{wu2015optimal}, the performance of randomized policy may be better than deterministic policy and easier to obtain. Therefore, we will find a randomized parametric event-based policy $\sigma$ for the proposed problem, i.e.,
\begin{equation}
\label{EQRandomParametricPolicy}
\sigma : \mathcal{E} \rightarrow \mathcal{P}(a_t^k)
\end{equation}
where $\mathcal{P}$ denotes a probability distribution over action space. Equation (\ref{EQRandomParametricPolicy}) means that the action $a_t^k$ will be chosen based on probability distribution $\mathcal{P}$ and observed event $e_t^k$. When an optimal randomized parametric event-based policy is obtained, the action with the highest probability can be selected for implementation in practice.

Secondly, after the microgrid operator controller allocates the charge ratio for each building, the charge controller in each building should decide which EV should be charged and keep the total number of charged EVs within $n_{t,m}^k + a_t^k (n_{t,c}^k - n_{t,m}^k)$. Therefore, we use a modified least-laxity-longer-processing-time-first (mLLLP) principle to select EVs to charge, which is introduced in \cite{jia2020structural}. The mLLLP principle generates a complete order among EVs and  selects EVs based on the remaining processing time $E_t^i/P$ and
the EV laxity $T_t^i-E_t^i/P$.

\subsection{Constrained Gradient-based Policy Optimization for EBO}
Due to the existence of constraint (\ref{EQTotalExchangeBound}), the proposed MDP is coupled among all the buildings in the microgrid. Therefore, in this paper we propose a constrained gradient-based policy optimization method to search the optimal randomized parametric event-based policy considering the coupled constraint.

First, we neglect constraint (\ref{EQTotalExchangeBound}) and the proposed model can be decoupled into single building optimization problem. For each building, let $\sigma$ and $\nu$ denotes two event-based policies, the following state-based performance difference formula can be derived based on \cite{Zhao2018PHD}.
\begin{equation}
\label{EQPerformanceDiff}
\begin{split}
& J_{t_0}^{k,\sigma}(s_{t_0}^k) - J_{t_0}^{k,\nu}(s_{t_0}^k) \\
& = \beta_{t_0}^{k,\sigma} P_{t_0}^{k,\sigma} (J_{t_0+1}^{k,\sigma} - J_{t_0+1}^{k,\nu}) \\
& + \beta_{t_0}^{k,\sigma}[c_{t_0}^{k,\sigma} + P_{t_0}^{k,\sigma} J_{t_0+1}^{k,\nu} - (c_{t_0}^{k,\nu} + P_{t_0}^{k,\nu} J_{t_0+1}^{k,\nu})] \\
& = \sum_{t=t_0}^{t_0+T_w-1}\beta_t^{k,\sigma}[c_{t}^{k,\sigma}-c_{t}^{k,\nu}+(P_t^{k,\sigma}-P_t^{k,\nu})J_{t+1}^{k,\nu})]
\end{split}
\end{equation}
where $J_{t_0}^{k,\sigma}$ and $J_{t_0}^{k,\nu}$ denotes the value function from stage $t_0$ to $t_0+T_w-1$ corresponding to the event-based policy $\sigma$ and $\nu$ for the $k$th building, $\beta_{t}^{k,\sigma}$ is the state distribution at stage $t$ corresponding to $\sigma$, $P_t^{k,\sigma}$ and $P_t^{k,\nu}$ denotes the transition probability at stage $t$ corresponding to $\sigma$ and $\nu$, $c_t^{k,\sigma}$ and $c_t^{k,\nu}$ denotes the one-step cost at stage $t$ corresponding to $\sigma$ and $\nu$. Note that the initial distribution $\beta_{t_0}^{k,\sigma}$ is independent with policy ${\sigma}$ or ${\nu}$. By using recursion and $\beta_{t_0}^{k,\sigma} P_{t_0}^{\sigma} = \beta_{t_0+1}^{k,\sigma}$, the last equation can be obtained.

Based on (\ref{EQPerformanceDiff}), we can extend the performance difference for event-based optimization, i.e.,
\begin{equation}
\label{EQEventsPerformanceDiff}
\begin{split}
& J_{t_0}^{k,\sigma}(s_{t_0}^k) - J_{t_0}^{k,\nu}(s_{t_0}^k) \\
& = \sum_{t=t_0}^{t_0+T_w-1} \sum_{s_{t}^k \in \mathcal{S}} \beta_t^{k,\sigma}(s_t) [c_{t}^{k,\sigma}-c_{t}^{k,\nu}+(P_t^{k,\sigma}-P_t^{k,\nu})J_{t_+1}^{k,\nu}] \\
& = \sum_{t=t_0}^{t_0+T_w-1} \sum_{s_{t}^k \in \mathcal{S}} \sum_{e_t^k \in \mathcal{E}} \beta_t^{k,\sigma}(s_t^k,e_t^k) \cdot \\
& \qquad \qquad \qquad [c_{t}^{k,\sigma}-c_{t}^{k,\nu}+(P_t^{k,\sigma}-P_t^{k,\nu})J_{t_+1}^{k,\nu}] \\
& = \sum_{t=t_0}^{t_0+T_w-1} \sum_{e_t^k \in \mathcal{E}} \beta_t^{k,\sigma}(e_t^k) \{ \sum_{s_{t}^k \in \mathcal{S}}  \beta_t^{k,\sigma}(s_t^k | e_t^k) \cdot \\
& \qquad \qquad \qquad [c_{t}^{k,\sigma}-c_{t}^{k,\nu}+(P_t^{k,\sigma}-P_t^{k,\nu})J_{t_+1}^{k,\nu}] \}
\end{split}
\end{equation}
where $\mathcal{S}$ denotes the state space. When policy $\nu$ is close to $\sigma$,  the performance gradient at stage $t_0$ can be derived as follows by observing event $e_{t_0}^k$,
\begin{equation}
\label{EQEventGradient}
\begin{split}
\frac{\partial J_{t_0}^{k,\sigma} (e_{t_0}^k)}{\partial \sigma_{t_0}} = & \beta_{t_0}^{k,\sigma_{t_0}} (e_{t_0}^k) \sum_{s_{t_0}^k \in \mathcal{S}} \beta_{t_0}^{k,\sigma_{t_0}}(s_{t_0}^k | e_{t_0}^k) \cdot \\
& [\frac{\partial c_{t_0}^{k,\sigma_{t_0}}}{\partial \sigma_{t_0}} + \frac{\partial P_{t_0}^{k,\sigma_{t_0}}}{\partial \sigma_{t_0}} J_{t_{0}+1}^{k,\sigma} ]
\end{split}
\end{equation}
where $\sigma_{t_0}$ denotes the detailed charge control policy at stage $t_{0}$ for the $k$th building.

Note that policy $\sigma$ is a randomized policy which selects the action with specific probability. Suppose there are $M$ actions for the $k$th building at stage $t_0$ and each action is denoted as $\alpha_{t_0}^{k,m}$.
Let $p_{t_0}^{k,m}$ denotes the probability to choose $\alpha_{t_0}^{k,m}$. Then there is the following relationship. The proof is given in the Appendix.
\begin{equation}
\label{EQTransferGradient}
\begin{split}
\frac{\partial P_{t_0}^{k,\sigma_{t_0}}}{\partial p_{t_0}^{k,m}} = & \frac{\sum_{i=1}^M p_{t_0}^{k,i} - p_{t_0}^{k,m}}{(\sum_{i=1}^M p_{t_0}^{k,i})^2} P(s_{{t_0}+1}^k|s_{t_0}^k,\alpha_{t_0}^{k,m}) \\
& + \sum_{i \neq m} \frac{-p_{t_0}^{k,i}}{(\sum_{i=1}^M p_{t_0}^{k,i})^2} P(s_{{t_0}+1}^k|s_{t_0}^k,\alpha_{t_0}^{k,i})
\end{split}
\end{equation}
\begin{equation}
\label{EQCostGradient}
\begin{split}
\frac{\partial c_{t_0}^{k,\sigma_{t_0}}}{\partial p_{t_0}^{k,m}} = & \frac{\sum_{i=1}^M p_{t_0}^{k,i} - p_{t_0}^{k,m}}{(\sum_{i=1}^M p_{t_0}^{k,i})^2} c(s_{t_0}^k,\alpha_{t_0}^{k,m}) \\
& + \sum_{i \neq m} \frac{-p_{t_0}^{k,i}}{(\sum_{i=1}^M p_{t_0}^{k,i})^2} c(s_{t_0}^k,\alpha_{t_0}^{k,i})
\end{split}
\end{equation}

Substituting (\ref{EQTransferGradient}) and (\ref{EQCostGradient}) into (\ref{EQEventGradient}), the policy gradient can be finally obtained which is shown below,
\begin{equation}
\label{EQEventProbGradient}
\begin{split}
\frac{\partial J_{t_0}^{k,\sigma} (e_{t_0}^k)}{\partial p_{t_0}^{k,m}} = & \beta_{t_0}^{k,\sigma_{t_0}} (e_{t_0}^k) \sum_{s_{t_0}^k \in \mathcal{S}} \beta_{t_0}^{k,\sigma_{t_0}}(s_{t_0}^k | e_{t_0}^k) \cdot \\
& \{ \frac{\sum_{i=1}^M p_{t_0}^{k,i} - p_{t_0}^{k,m}}{(\sum_{i=1}^M p_{t_0}^{k,i})^2} [c(s_{t_0}^k,\alpha_{t_0}^{k,m}) + V(s_{t_0}^k,\alpha_{t_0}^{k,m})] \\
& + \sum_{i \neq m} \frac{-p_{t_0}^{k,i}}{(\sum_{i=1}^M p_{t_0}^{k,i})^2} [c(s_{t_0}^k,\alpha_{t_0}^{k,i}) + V(s_{t_0}^k,\alpha_{t_0}^{k,i})]\}
\end{split}
\end{equation}
where
\begin{equation}
V(s_{t_0}^k,\alpha_{t_0}^{k,m}) = \sum_{s_{t_0+1}^k \in \mathcal{S}}  P(s_{t_0+1}^k | s_{t_0}^k,\alpha_{t_0}^{k,m}) J_{t_0+1}^{k,\sigma}(s_{t_0+1}^k)
\end{equation}
denotes the incurred future total cost when taking action $\alpha_{t_0}^{k,m}$ for current state $s_{t_0}^k$ and $\frac{\partial J_{t_0}^{k,\sigma} (e_{t_0}^k)}{\partial \sigma_{t_0}} = (\frac{\partial J_{t_0}^{k,\sigma} (e_{t_0}^k)}{\partial p_{t_0}^{k,1}},\frac{\partial J_{t_0}^{k,\sigma} (e_{t_0}^k)}{\partial p_{t_0}^{k,2}},...,\frac{\partial J_{t_0}^{k,\sigma} (e_{t_0}^k)}{\partial p_{t_0}^{k,M}})$.

Based on the policy gradient (\ref{EQEventProbGradient}), the randomized parametric event-based policy can be updated as follows during policy optimization,
\begin{equation}
\label{EQPolicyUpdate}
\sigma_{t_0,j+1}  = \sigma_{t_0,j} - \delta_j \frac{\partial J_{t_0}^{k,\sigma_{j}} (e_{t_0}^k)}{\partial \sigma_{t_0}^{k}}
\end{equation}
where $\sigma_{t_0,j}$ denotes the updated event-based policy at $j$th iteration, $\delta_j = 1/(1+\xi j)$ denotes the update step at $j$th iteration and $\xi$ denotes the decay factor.

Due to the uncertainties in the DRE generation and EV charging demand, it is impractical to analytically compute equation (\ref{EQEventProbGradient}) under expectation. Therefore, the Monte Carlo simulation method is adopted to estimate (\ref{EQEventProbGradient}). The estimation algorithm is summarized in Algorithm 1.

\begin{algorithm}[h]
\caption{Policy Gradient Estimation}
\begin{algorithmic}[1]
\STATE Input:policy $\sigma$.
\STATE Generate and record sample paths under policy $\sigma$, i.e., $\{ s_{t_0,i}^k, a_{t_0,i}^k, s_{t_0+1,i}^k, a_{t_0+1,i}^k,...,s_{t_0+T_w-1,i}^k, a_{t_0+T_w-1,i}^k \}$ where $i=1,2,...,L$, $L$ is the total number of sample paths, $s_{t_0,i}^k$ and $a_{t_0,i}^k$ denotes the observed state and selected action of the $k$th building in the $i$th sample path.
\STATE Compute the number of occurrences $L(e_{t_0}^k)$ when event $e_{t_0}^k$ happens. Then there is $\beta_{t_0}^{k,\sigma_{t_0}} = L(e_{t_0}^k)/L$.
\STATE Compute the number of occurrences $L(s_{t_0}^k | e_{t_0}^k)$ when event $e_{t_0}^k$ happens and the system observes state $s_{t_0}^k$. Then there is $\beta_{t_0}^{k,\sigma_{t_0}}(s_{t_0}^k | e_{t_0}^k) = L(s_{t_0}^k | e_{t_0}^k)/L(e_{t_0}^k)$.
\STATE Compute the number of occurrences $L(e_{t_0}^k,s_{t_0}^k,\alpha_{t_0}^{k,m})$ when $e_{t_0}^k$ happens and state-action pair $(s_{t_0}^k,\alpha_{t_0}^{k,m})$ appears (Denote this sample path set as $\mathcal{L}(e_{t_0}^k,s_{t_0}^k,\alpha_{t_0}^{k,m})$). Then there is $V(s_{t_0}^k,\alpha_{t_0}^{k,m}) = \frac{1}{L(e_{t_0}^k,s_{t_0}^k,\alpha_{t_0}^{k,m})}\sum_{i \in \mathcal{L}(e_{t_0}^k,s_{t_0}^k,\alpha_{t_0}^{k,m})} \sum_{t=t_0+1}^{t_0+T_w-1} c_{t}^{k,\sigma}$.
\STATE Substituting the above equations into (\ref{EQEventProbGradient}).
\end{algorithmic}
\end{algorithm}

As mentioned before, the derived policy gradient neglects the coupled constraint (\ref{EQTotalExchangeBound}). In order to satisfy this transmission safety constraint, the following adjusting mechanism is proposed to ensure the feasibility of policy $\sigma$.

\textit{Adjusting Step I:} If the total exchange power exceed the upper bound of (\ref{EQTotalExchangeBound}) by $\Delta$, the total number of EVs to be charged should be reduced by $\Delta/P$. Each building should reduce the number of charged EVs by $(\partial J_{t_0}^{k,\sigma}  \Delta /\partial \sigma_{t_0}^{k})/(P \sum_{k=1}^K \partial J_{t_0}^{k,\sigma} /\partial \sigma_{t_0}^{k})$. The policy $\sigma$ can be updated by solving the following Quadratic Programming problem,
\begin{equation}
\label{EQAdjustUpper}
\begin{split}
& \quad \min_{p_{t_0, j+1}^{k,i} \in [0,1]} \sum_{i=1}^M (p_{t_0, j+1}^{k,i} - p_{t_0,j}^{k,i})^2 \\
& \textit{s.t.} \\
& \sum_{i=1}^M p_{t_0, j+1}^{k,i} \alpha_{t_0}^{k,m} = \sum_{i=1}^M p_{t_0, j}^{k,i} \alpha_{t_0}^{k,m} - \\
& \qquad \qquad \qquad \quad \frac{\partial J_{t_0}^{k,\sigma}  / \partial \sigma_{t_0}^{k}}{ \sum_{k=1}^K \partial J_{t_0}^{k,\sigma} /\partial \sigma_{t_0}^{k}} \cdot \frac{\Delta }{P(n_{t,c}^k - n_{t,m}^k)} \\
\end{split}
\end{equation}
where $p_{t_0, j+1}^{k,i}$ denotes the selection probability in $\sigma_{t_0, j+1}$.

\textit{Adjusting Step II:} If the total exchange power is below the lower bound of (\ref{EQTotalExchangeBound}) with $-\Delta$, the total number of EVs to be charged should be increased by $\Delta/P$. Each building should increase the number of charged EVs by $(1-\partial J_{t_0}^{k,\sigma} /\partial \sigma_{t_0}^{k} / \sum_{k=1}^K \partial J_{t_0}^{k,\sigma} /\partial \sigma_{t_0}^{k}) \Delta  /  P $. Similarly, the policy $\sigma$ can be updated by solving the following problem,
\begin{equation}
\label{EQAdjustLower}
\begin{split}
& \quad \min_{p_{t_0, j+1}^{k,i} \in [0,1]} \sum_{i=1}^M (p_{t_0, j+1}^{k,i} - p_{t_0,j}^{k,i})^2 \\
& \textit{s.t.} \\
& \sum_{i=1}^M p_{t_0, j+1}^{k,i} \alpha_{t_0}^{k,m} = \sum_{i=1}^M p_{t_0, j}^{k,i} \alpha_{t_0}^{k,m} + \\
& \qquad \qquad \quad  (1-\frac{\partial J_{t_0}^{k,\sigma} / \partial \sigma_{t_0}^{k}}{\sum_{k=1}^K \partial J_{t_0}^{k,\sigma} /\partial \sigma_{t_0}^{k}}) \cdot \frac{\Delta }{P(n_{t,c}^k - n_{t,m}^k)} \\
\end{split}
\end{equation}

Note that as $\partial J_{t_0}^{k,\sigma} /\partial \sigma_{t_0}^{k}$ can be considered as the marginal operation cost for the $k$th building, the proposed adjusting mechanism allocates the reduced or increased number of charged EVs for each building based on this marginal cost. When required to reduce the charge demand, the building with the large marginal cost should largely reduce its charged EVs. On the contrary, when required to increase the charge demand, the building with small marginal cost should largely increase its EVs to be charged. The motivation of solving (\ref{EQAdjustUpper}) and (\ref{EQAdjustLower}) is to minimize the probability difference between the adjacent policy $\sigma_{t_0,j+1}$ and $\sigma_{t_0,j}$ while reduce or increase the expected charge ratio to satisfy the allocated reduced or increased charge demand for the $k$th building.

\begin{figure}
\centering
\setlength{\abovecaptionskip}{-6pt}
\includegraphics[height=0.18\textheight]{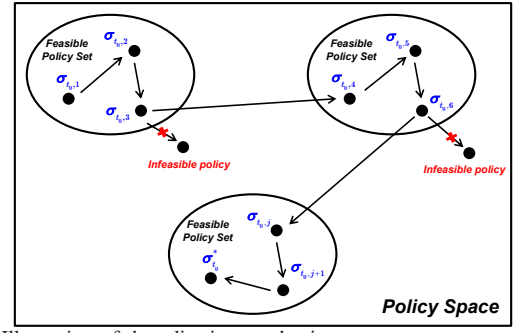}
\caption{Illustration of the adjusting mechanism.}
\label{AdjustIllustrate}
\end{figure}

The idea of this adjust mechanism can be denoted as Fig. \ref{AdjustIllustrate}. As there exists large number of discrete variables and non-linear constraints, the feasible policy space can be considered as disconnected. When constraint (\ref{EQTotalExchangeBound}) is not violated, the policy update happens within a feasible policy set. When violated, the policy update should transfer to another feasible policy set based on this adjust mechanism. In the end, the proposed constrained gradient-based policy optimization for the problem is summarized in Algorithm 2.

\begin{algorithm}[h]
\caption{Constrained Gradient-based Policy Optimization}
\begin{algorithmic}[1]
\FOR{$t_0=1,2,...,T$}
\STATE Set $j \rightarrow 0$ and select the initial policy $\sigma_{t_0,j}$.
\FOR{k=1,2,...,K}
\STATE Compute the policy gradient based on Algorithm 1 when observing the event $e_{t_0}^k$.
\ENDFOR
\STATE Check whether constraint (\ref{EQTotalExchangeBound}) is violated. If violated, update policy by adjusting mechanism. Go to Step 3.
\STATE If $|| \frac{\partial J_{t_0}^{k,\sigma_j} }{\partial \sigma_{t_0}} ||_2 \leq \epsilon$ or $||
\frac{\partial J_{t_0}^{k,\sigma_{j}}}{\partial \sigma_{t_0}} - \frac{\partial J_{t_0}^{k,\sigma_{j-1}}}{\partial \sigma_{t_0}} ||_2 \leq \epsilon$, go to Step 1.
\STATE  Update policy using (\ref{EQPolicyUpdate}) and go to Step 3.
\ENDFOR
\end{algorithmic}
\end{algorithm}

\section{Numerical Results}
In this section, we evaluate the proposed method for EV charging scheduling via simulations. The charge control policy of different types of buildings, the performance and comparison of the proposed solution method are analyzed in the following experiments.

\subsection{Parameter Settings}
\begin{figure}
\centering
\setlength{\abovecaptionskip}{-6pt}
\includegraphics[height=0.2\textheight]{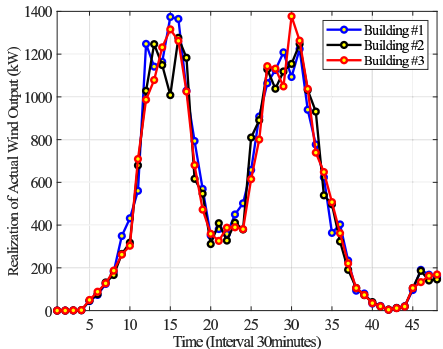}
\caption{Realization of actual wind output in three buildings.}
\label{WindPower}
\end{figure}

We take building load data from \cite{wang2015integrated} as shown in Table \ref{TBLoadWind}. In the experiment, we consider there are three buildings in the microgrid. The first two are residential buildings and the rest is office building. We take distributed wind generation data from \cite{huang2016multi} and set as the predicted value. The actual output of DRE in each building is assumed to follow normal distribution with predicted value as the mean value and $10\%$ of the predicted value as the standard deviation. Fig. \ref{WindPower} shows the realization of the actual output of DRE in three buildings. The time-of-use electricity price is shown in Table \ref{TBEP}.

\begin{table}[!t]
\scriptsize\centering
\caption{Building Load Data (unit: kW)}
\label{TBLoadWind}
\begin{tabular}{c c c c|c c c c}
\toprule
Time(h) & $k=1$ & $k=2$ & $k=3$ & Time(h) &  $k=1$ & $k=2$ & $k=3$ \\
\midrule
1 & 139 & 180 & 367 & 13 & 185 & 216 & 953 \\
2 & 103 & 120 & 353 & 14 & 420 & 516 & 953 \\
3 & 144 & 180 & 333 & 15 & 430 & 516 & 947 \\
4 & 127 & 147 & 333 & 16 & 743 & 876 & 967 \\
5 & 151 & 180 & 433 & 17 & 1132 & 1356 & 953 \\
6 & 150 & 180 & 387 & 18 & 1340 & 1596 & 1053 \\
7 & 67 & 84 & 520 & 19 & 2681 & 2160 & 1033 \\
8 & 202 & 240 & 567 & 20 & 1780 & 1643 & 1000 \\
9 & 216 & 264 & 820 & 21 & 1760 & 1400 & 967 \\
10 & 151 & 191 & 1053 & 22 & 1648 & 1320 & 820 \\
11 & 147 & 170 & 967 & 23 & 1251 & 1500 & 700 \\
12 & 150 & 258 & 973 & 24 & 559 & 660 & 593 \\
\bottomrule
\end{tabular}
\end{table}

\begin{table}[!t]
\scriptsize
\centering
\caption{\label{TBEP}Electricity Price $\omega_t$ }
\begin{tabular}{c c}
\toprule
Price & Time\\
\midrule
0.3515RMB/kWh & 23:00-6:00\\
0.8135RMB/kWh & 7:00-10:00\\
0.4883RMB/kWh & 11:00-18:00\\
0.8135RMB/kWh & 19:00-22:00\\
\bottomrule
\end{tabular}
\end{table}

We consider there are 200 EVs in the microgrid and their drivers live evenly in the two residential buildings and work in the office building. In the experiment, we assume that the departure time from residential and office buildings follow normal distribution $\mathcal{N}(7:00,60\text{min})$ and $\mathcal{N}(17:00,60\text{min})$, respectively. The trip time between building \#1 (k=1) and building \#3 (k=3) follows normal distribution $\mathcal{N}(60\text{min},30\text{min})$. The trip time between building \#2 (k=2) and building \#3 (k=3) follows normal distribution $\mathcal{N}(90\text{min},30\text{min})$.

The battery specification of the Nissan Leaf EV \cite{WebsiteNissan} is used in the experiments. The required charging energy $\eta_{t+1}^i$ of future parking event is sampled based on the probability distribution of the trip distance and the electric drive efficiency which is introduced in \cite{huang2016multi}. We take the parameters of HES from \cite{wu2021hydrogen}. The detailed parameters are shown in Table \ref{TBPara}.

\begin{table}[!t]
\scriptsize
\centering
\caption{\label{TBPara}Parameter Settings}
\begin{tabular}{c c c c}
\toprule
Parameter & Setting & Parameter & Setting\\
\midrule
$E_{\text{cap}}$ & 36kWh & $P$ & 3.6kW\\
$\psi^c$ & 0.92 & $\kappa_e^{\text{cap}}$ & 166.65kWh\\
$h^{\text{cap}}$ & 50kW & $\eta^c$ & 0.82\\
$\eta^{dc}$ & 0.62 & $\epsilon$ & 0.1\\
$ \underline{G}$  & -5600kW & $\overline{G}$ & 5600kW\\
$L$ & 50 &  $\xi$ & 0.1\\
\bottomrule
\end{tabular}
\end{table}

In the experiment, we consider this problem on a daily basis with $T=48$, $\Delta T = 30$minutes and $T_w=12$. The event is evenly discretized by 0.1. The action selection probabilities of the initial policy $\sigma$ are all set as equal.

\subsection{Result Analysis}

\begin{figure*}
  \centering
  \subfigure[Optimized Selection Probability ($k=1$)]{
    \label{fig:subfig:a} 
    \includegraphics[width=0.32\textwidth]{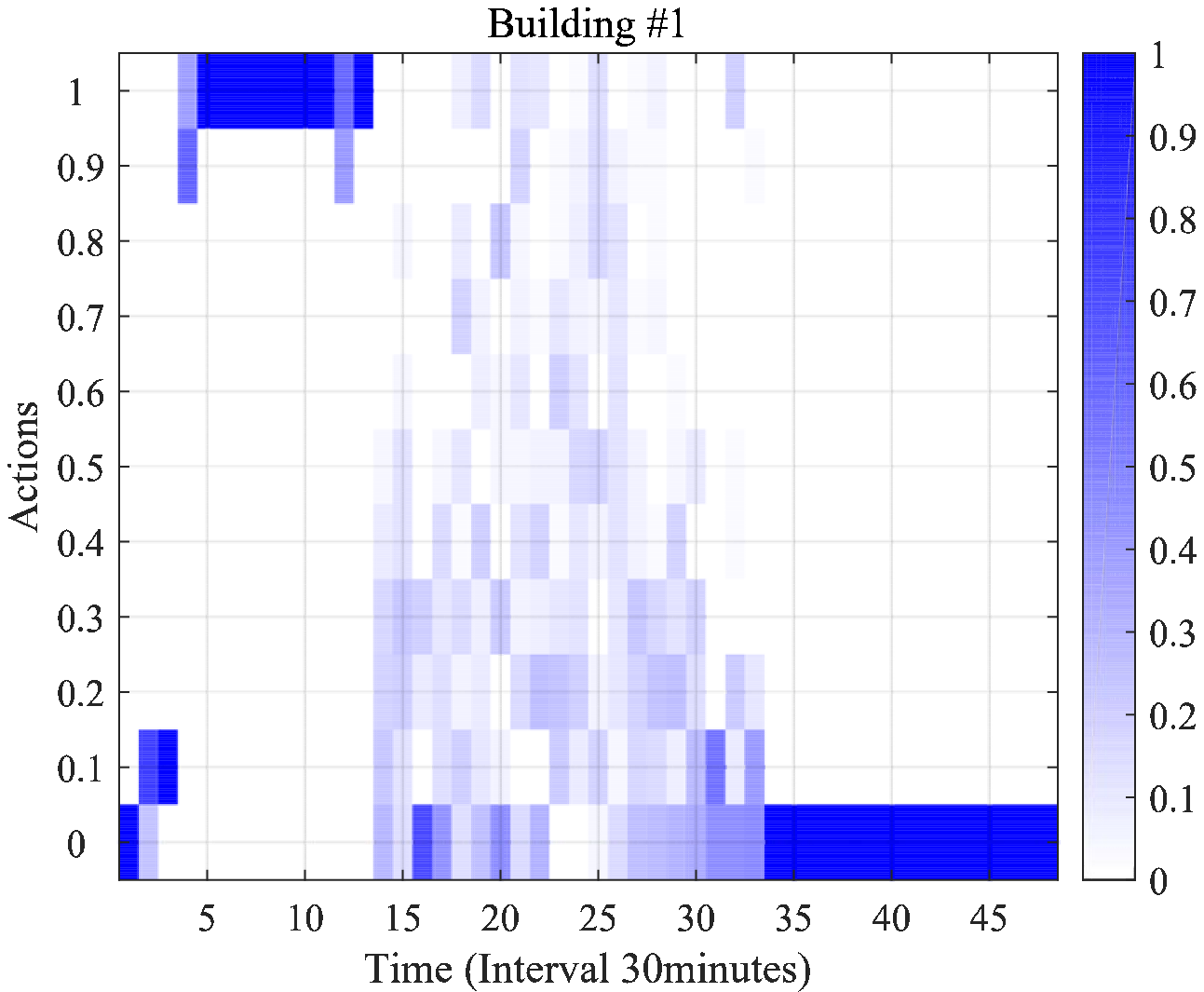}}
  \subfigure[Optimized Selection Probability ($k=2$)]{
    \label{fig:subfig:b} 
    \includegraphics[width=0.32\textwidth]{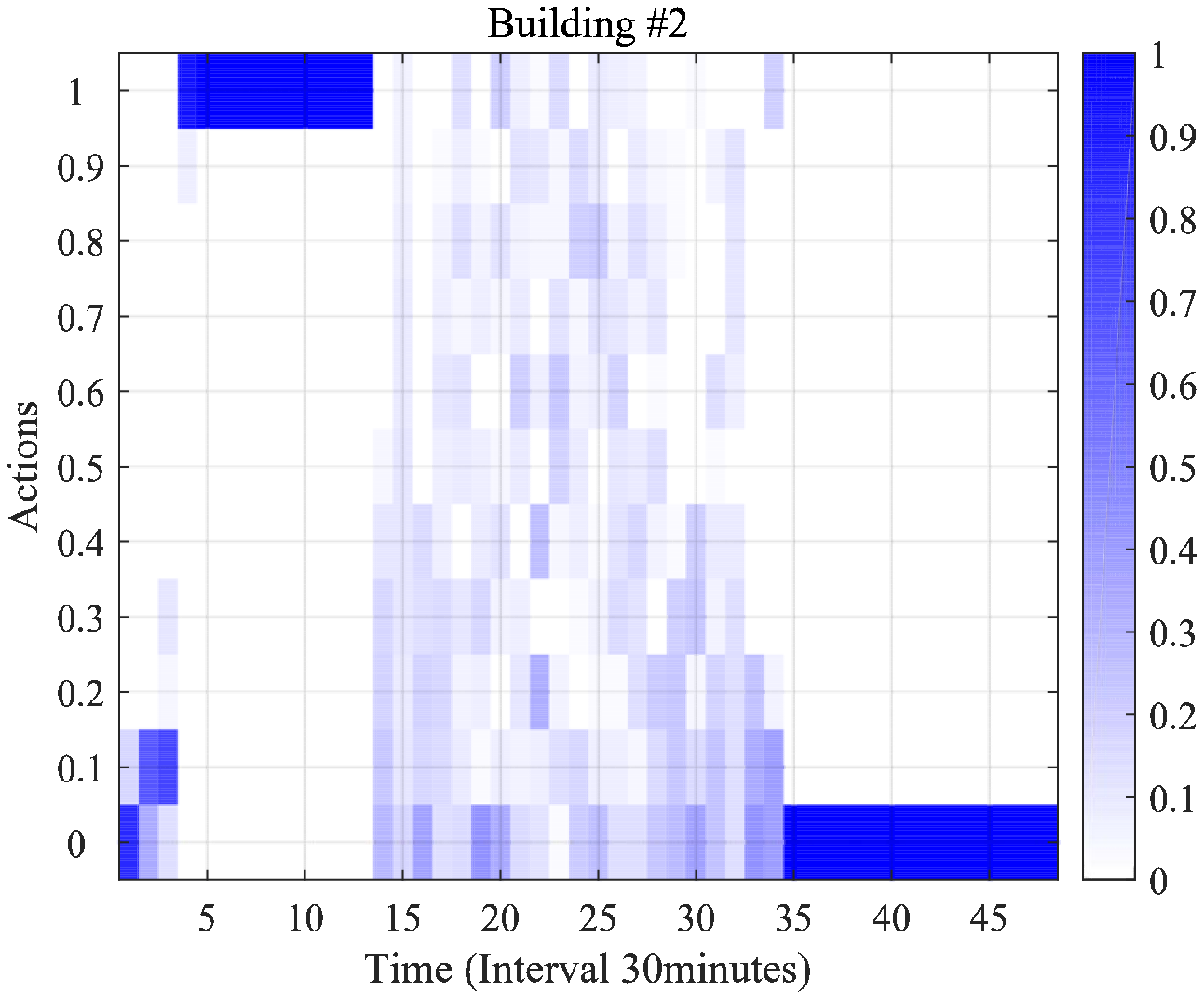}}
  \subfigure[Optimized Selection Probability ($k=3$)]{
    \label{fig:subfig:c} 
    \includegraphics[width=0.32\textwidth]{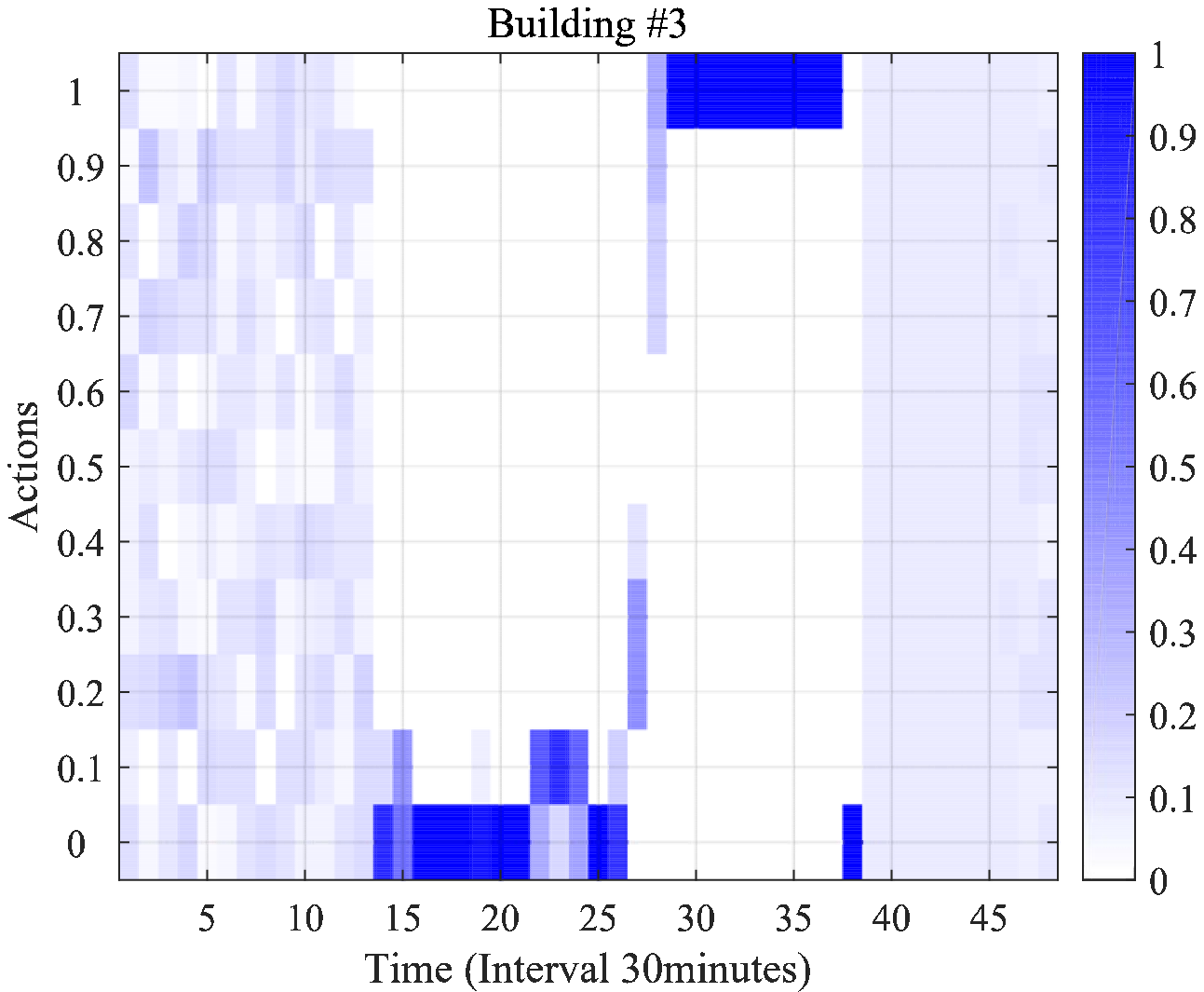}}

    \subfigure[Total Charge Power and Event ($k=1)$]{
    \label{fig:subfig:d} 
    \includegraphics[width=0.32\textwidth]{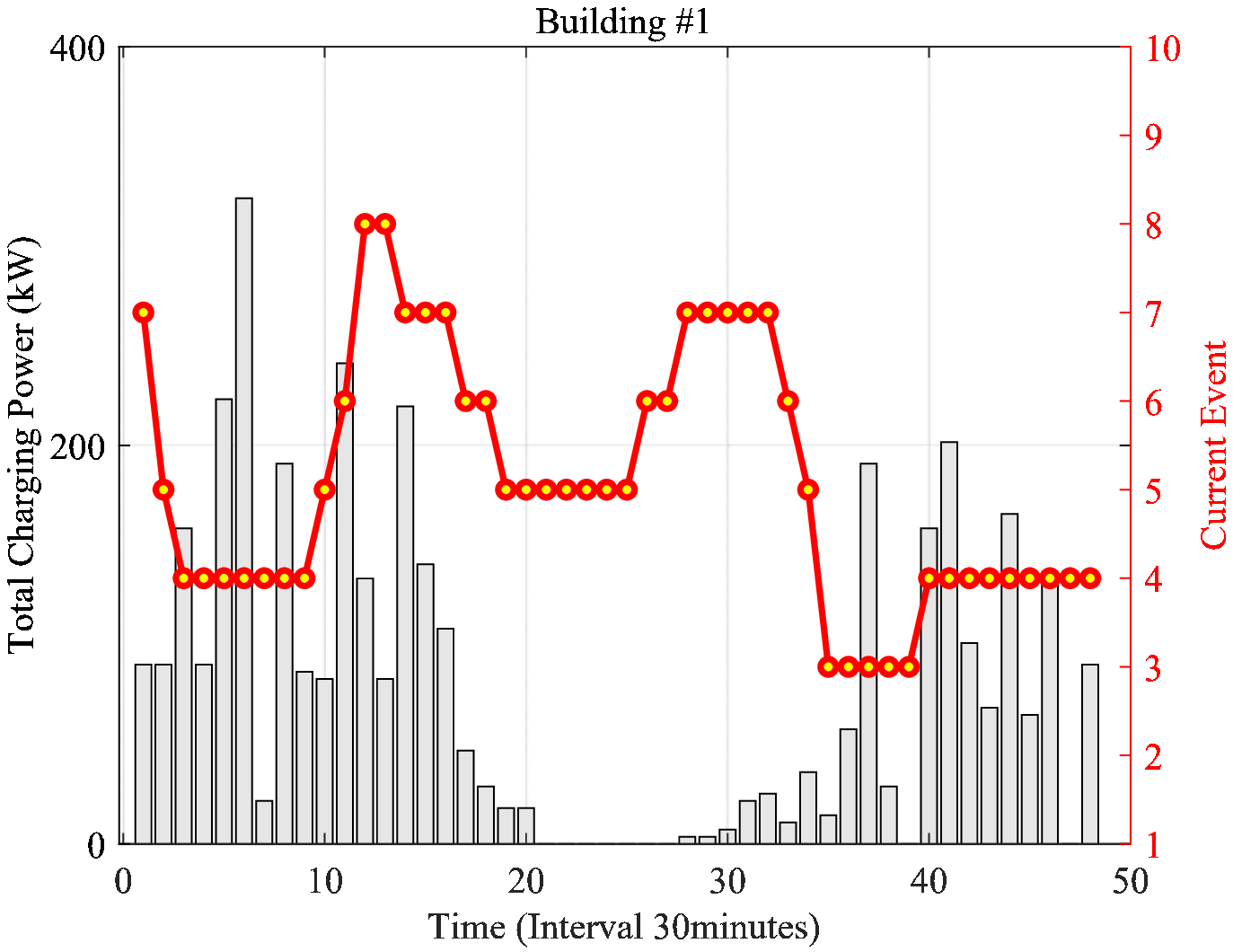}}
  \subfigure[Total Charge Power and Event ($k=2)$]{
    \label{fig:subfig:e} 
    \includegraphics[width=0.32\textwidth]{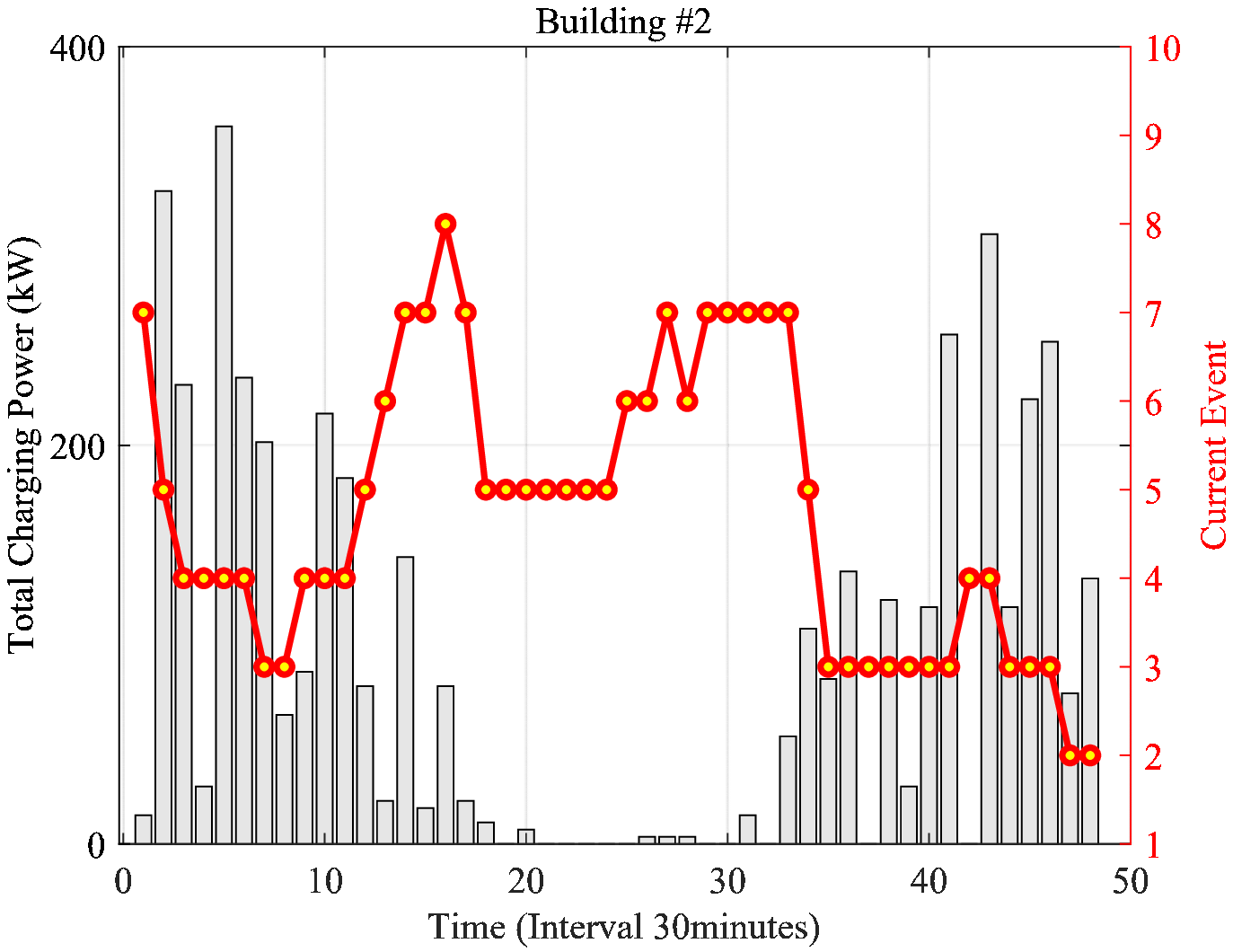}}
  \subfigure[Total Charge Power and Event ($k=3)$]{
    \label{fig:subfig:f} 
    \includegraphics[width=0.32\textwidth]{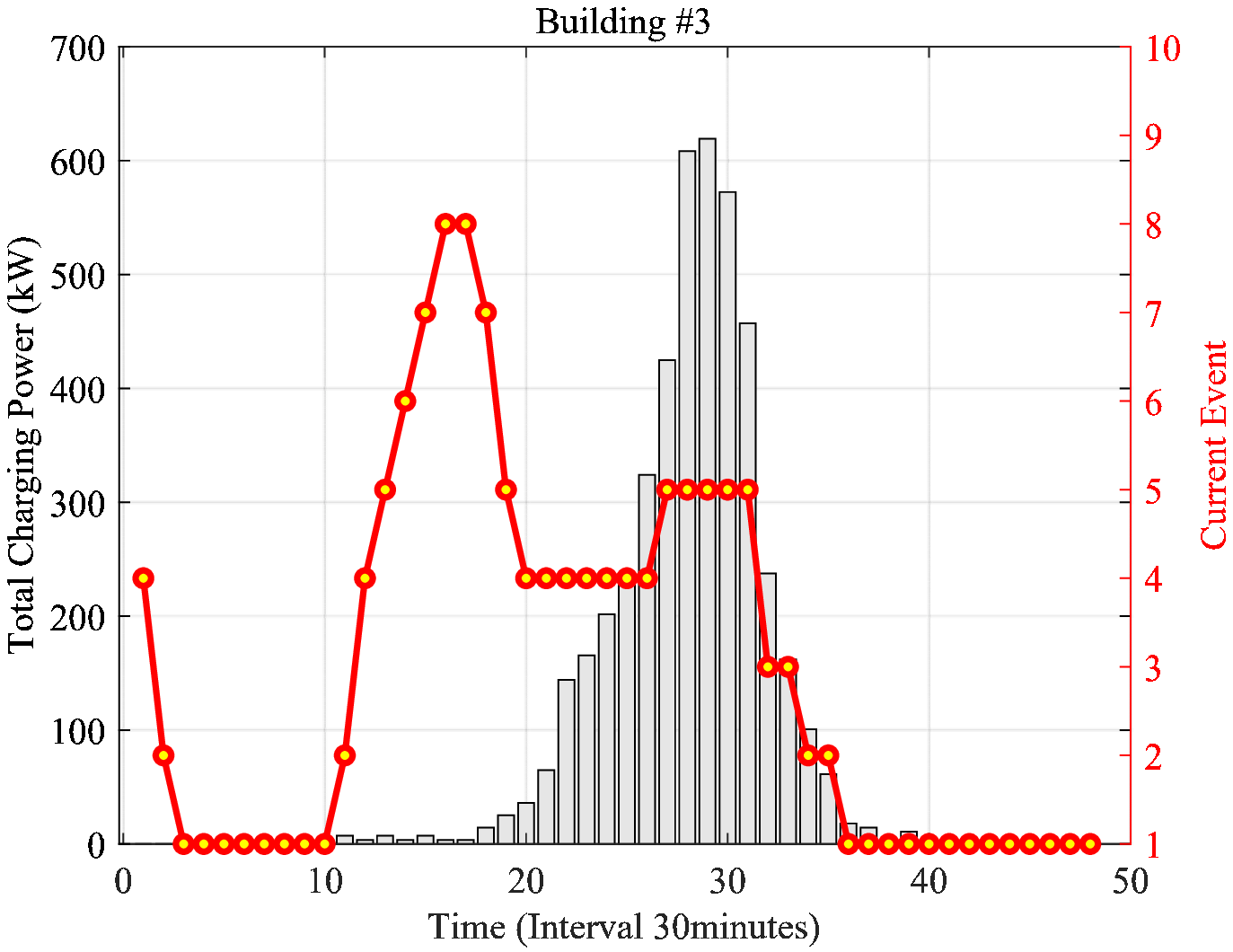}}

  \subfigure[SOC of HES ($k=1$)]{
    \label{fig:subfig:g} 
    \includegraphics[width=0.32\textwidth]{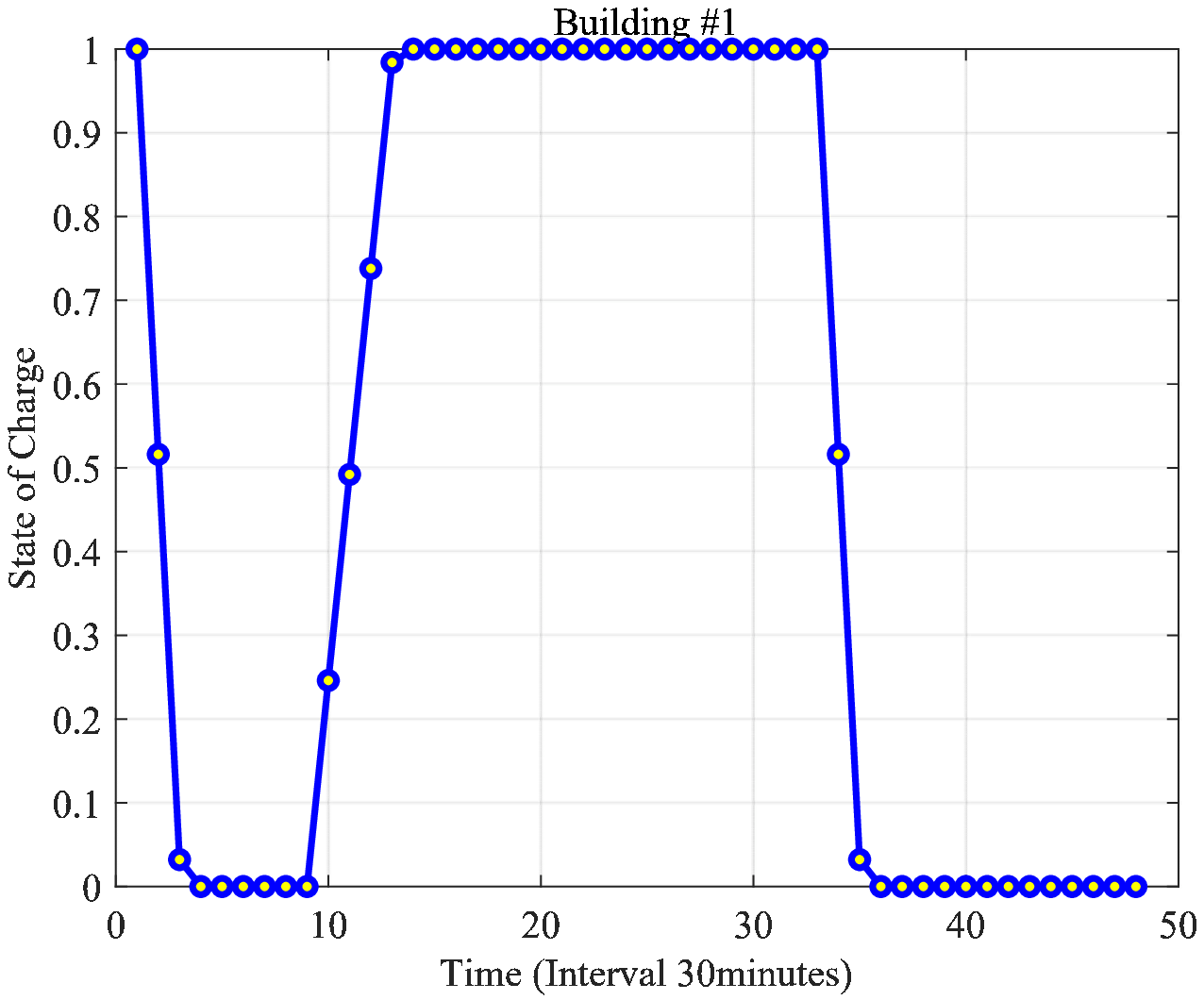}}
  \subfigure[SOC of HES ($k=2$)]{
    \label{fig:subfig:h} 
    \includegraphics[width=0.32\textwidth]{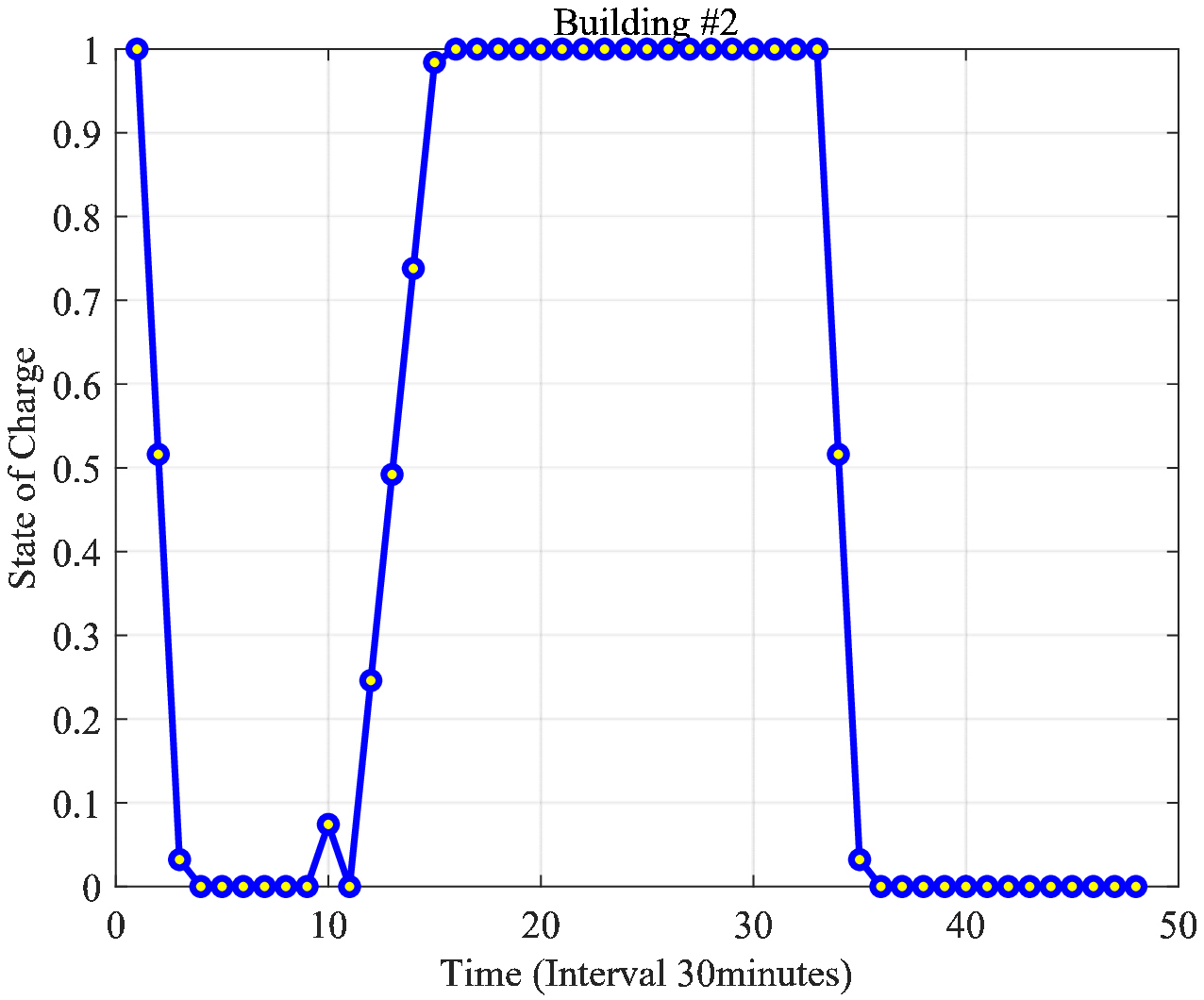}}
  \subfigure[SOC of HES ($k=3$)]{
    \label{fig:subfig:i} 
    \includegraphics[width=0.32\textwidth]{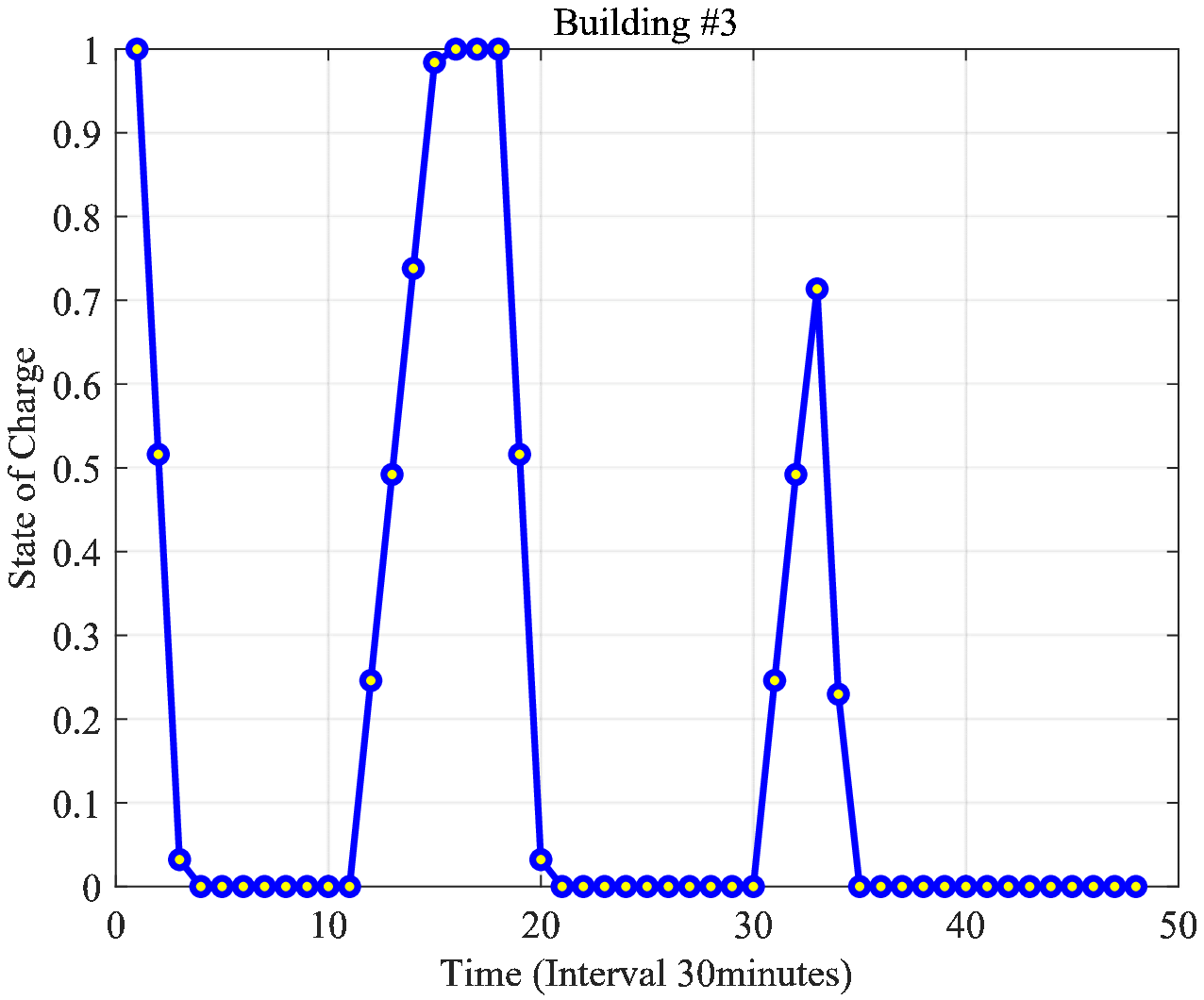}}
  \caption{Scheduling results for a microgrid of three buildings with 200 EVs}
  \label{fig:subfig} 
\end{figure*}

The scheduling results are shown in Fig. \ref{fig:subfig} which shows the optimized selection probability, total charging power, observed event and SOC of HES for three buildings at each stage. Note that Fig. \ref{fig:subfig:a}, Fig. \ref{fig:subfig:b} and Fig. \ref{fig:subfig:c} shows the optimized selection probability of each charge ratio action in the randomized event-based policy after observing the event as shown in Fig. \ref{fig:subfig:d}, Fig. \ref{fig:subfig:e} and Fig. \ref{fig:subfig:f}. The darker color indicates higher selection probability. From Fig. \ref{fig:subfig:a}, Fig. \ref{fig:subfig:b} and Fig. \ref{fig:subfig:c}, it can be seen that the selection probabilities tend to achieve high value near the departure time, such as 7:00 in building \#1 and building \#2 and 17:00 in building \#3. The reason why the charging probability is small at the arrival and becomes large at the departure lies in two aspects. The first is that parking deadline approaches and the charging demand should be satisfied. Another important reason is that the distributed wind power begins to increase during time interval (2:30-7:30) and (14:00-16:30). Note that as there are few EVs in building \#1 and building \#2 during (8:00-16:00), there is no action whose selection probability is significantly higher than others. The same reason holds for building \#3 during (24:00-6:00). Furthermore, as there is no EV parked in building \#3 after 20:00 considering time window $T_w$ set as 6 hours, the policy gradient keeps zero and the selection probability for each action remains unchanged and equal.

The total charging power and observed event is shown in Fig. \ref{fig:subfig:d}, Fig. \ref{fig:subfig:e} and Fig. \ref{fig:subfig:f}. Due to the stochastic charging demand of EVs and their distinct departure, the charging behavior occurs during (16:00-10:00) for building \#1 and building \#2 and during (10:00-18:00) for building \#3. Furthermore, it can be found that the peak of the total charging power occurs later in building \#2 than building \#1 after 15:00. This is because that the trip time from building \#3 to building \#2 is longer than that from building \#3 to building \#1. Based on the delayed feature of the optimized selection probability in building \#3, the peak occurrence of charging power are also delayed to 15:00 and gradually decrease due to the departure of EVs. It can also be found that the trend of the observed event is similar with the trend of the SOC of HES in each building as shown in Fig. \ref{fig:subfig:g}, Fig. \ref{fig:subfig:h} and Fig. \ref{fig:subfig:i}. This is because the SOC of HES is one of the main factor which influence the value of the observed event. The difference between the trend of observed event and SOC of HES is caused by the distributed wind power generation and EV charging elasticity. In Fig. \ref{fig:subfig:g}, Fig. \ref{fig:subfig:h} and Fig. \ref{fig:subfig:i}, the decreasing of SOC indicates that the HES provides power for balancing building load and EV charging load and the increasing of SOC indicates that the excess generation of distributed wind power. The small peak at 10:00 in building \#2 is because the insufficient generation of distributed wind power and larger load demand comparing with building \#1.

In order to analyze the performance of the optimized event-based EV charging policy, we compare the derived event-based policy with rule-based charging policy and ideal charging policy. The rule-based charging policy will satisfy the EV charging demand as soon as possible once connected to the charging pile in the building. The ideal charging policy is derived by implementing model predictive control (MPC) method with precise information of the EV charging demand and wind power generation and the same length of sliding window. Particularly, the optimal scheduling of HES are also considered as the control variables in MPC. In this way, it will introduce some non-linear constraints in the MPC model which should be linearized, such as the product between integer variables and continuous variables. The performance of the above three policies are shown in Table \ref{TBCOST}. It can be seen that the rule-based policy achieves highest operation cost as the EV charging control has no relationship with the building load and supply. On the contrary, the ideal charging achieves lowest operation cost. However, this policy can not be implemented in practice due to its requirement of seeing the future. Compared the event-based policy with ideal charging, it can be found that the performance of our policy is close to the idea policy and better than the rule-based charging policy. This demonstrates the effectiveness of the proposed EV charging control method.

\begin{table}[!t]
\scriptsize
\centering
\caption{\label{TBCOST}Total Operation Cost of Microgrid under Different Policies}
\begin{tabular}{c c c c}
\toprule
Rule-based Charging & Event-based Charging  & Ideal Charging \\
\midrule
28163RMB & 24965RMB	& 24811RMB	\\
\bottomrule
\end{tabular}
\end{table}

\begin{figure}
\centering
\setlength{\abovecaptionskip}{-6pt}
\includegraphics[height=0.2\textheight]{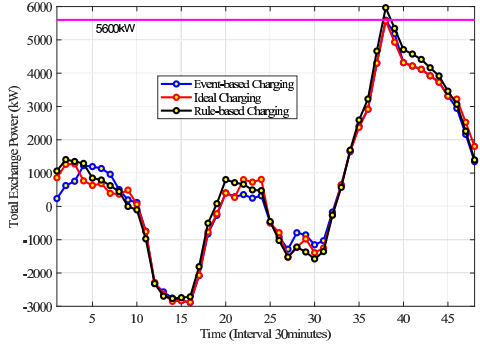}
\caption{Total exchange power of microgrid under different policies.}
\label{TotalExchangePower}
\end{figure}

The total exchange power of microgrid under the above three policies are also shown in Fig. \ref{TotalExchangePower}. The peak of the total exchange power at 19:00 is caused by the high building load in building \#1 and building \#2 at night. It can be seen that the rule-based charging policy exceed the maximum transmission power $\overline{G}=5600\text{kW}$ while the peak of the latter two policies are below this upper bound. This indicates that the proposed constrained gradient-based policy optimization method can ensure the transmission safety. It can be seen from the figure that the main reason why the event-based charging and ideal charging outperform the rule-based charging lies in two aspects. The first is the total exchange power of the event-based policy and ideal policy postpones during (9:00-12:00) to enjoy low electricity price after 11:00. The second is the ahead of schedule of the event-based policy and ideal policy during (13:00-22:00) to avoid high electricity price during (19:00-22:00) and enjoy free wind power generation around 15:00.

\begin{figure}
\centering
\setlength{\abovecaptionskip}{-6pt}
\includegraphics[height=0.2\textheight]{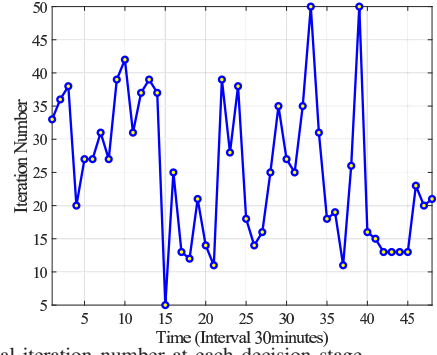}
\caption{Total iteration number at each decision stage.}
\label{IterationNumber}
\end{figure}

Lastly, it is important to investigate the convergence rate of the proposed constrained gradient-based policy optimization. Therefore, Fig. \ref{IterationNumber} shows the total iteration number at each decision stage. For this experiment, the minimum iteration number is 5 which happens at 7:30 and the maximum iteration number is 50 which happens at 16:30 and 19:30. It takes about 25.3 iterations in average to find the optimal event-based charging policy at each decision stage. The average computation time at each decision stage is 5.2 minutes under the simulation environment i7-11700K@3.60GHz. This indicates the proposed algorithm requires a handful of iterations and its running time can be acceptable in practice.

\section{Conclusion}
In this paper, the EV charging scheduling problem in a microgrid of buildings is studied to optimize the total operation cost of the microgrid while ensuring its transmission safety. The MDP formulation is introduced to represent the uncertain supply and EV charging demand in the buildings. In order to alleviate the large state and action space difficulties, we reformulated it within a event-based optimization framework with searchable control policy space. A constrained gradient-based policy optimization approach is proposed to find an optimal randomized parametric control policy for EV charging. We analyze the structure of the control policy through numerical experiments and demonstrate the proposed method can reduce the total operation cost while ensuring transmission safety in the microgrid of buildings.


%

\appendices
\section{Proof of Equation (\ref{EQTransferGradient}) and (\ref{EQCostGradient})}
\begin{proof}
Based on the randomized control policy $\sigma$, there is
\begin{equation}
P(\alpha_{t_0}^{k,m}|s_{t_0}^k) = \frac{p_{t_0}^{k,m}}{\sum_{i=1}^M p_{t_0}^{k,i}}
\end{equation}
\begin{equation}
\label{EQTransferProb}
P_{t_0}^{k,\sigma}(s_{t_0+1}^k|s_{t_0}^k) = \sum_{i=1}^M P(\alpha_{t_0}^{k,i}|s_{t_0}^k) P(s_{t_0+1}^k | s_{t_0}^k,\alpha_{t_0}^{k,i})
\end{equation}

For the selection probability $P(\alpha_{t_0}^{k,m}|s_{t_0}^k)$, there is
\begin{equation}
\label{EQSelectionGradient}
\frac{\partial P(\alpha_{t_0}^{k,i}|s_{t_0}^k)}{\partial p_{t_0}^{k,m}} =
\begin{cases}
\frac{\sum_{i=1}^M p_{t_0}^{k,i} - p_{t_0}^{k,m}}{(\sum_{i=1}^M p_{t_0}^{k,i})^2}, & \text{if} \ i = m\\
\frac{-p_{t_0}^{k,i}}{(\sum_{i=1}^M p_{t_0}^{k,i})^2}, & \text{if} \ i \neq m
\end{cases}
\end{equation}
. As only $P(\alpha_{t_0}^{k,i}|s_{t_0}^k)$ depends on $p_{t_0}^{k,m}$ in equation (\ref{EQTransferProb}), equation (\ref{EQTransferGradient}) is obtained by taking derivative of equation (\ref{EQTransferProb}) with $p_{t_0}^{k,m}$ and substituting (\ref{EQSelectionGradient}) into it.

Similarly, for the one-step cost $c_{t_0}^{k,\sigma_{t_0}}$, there is
\begin{equation}
\label{EQCostRelation}
c_{t_0}^{k,\sigma}(s_{t_0}^k) = \sum_{i=1}^M P(\alpha_{t_0}^{k,i}|s_{t_0}^k) c(s_{t_0}^k,\alpha_{t_0}^{k,i})
\end{equation}
. Taking derivative of equation (\ref{EQCostRelation}) with $p_{t_0}^{k,m}$ and substituting (\ref{EQSelectionGradient}) into it, equation (\ref{EQCostGradient}) can be obtained.
\end{proof}


%
%

\ifCLASSOPTIONcaptionsoff
  \newpage
\fi

\bibliographystyle{IEEEtran}
\bibliography{IEEEabrv,myref}

\end{document}